\documentclass[11pt]{article}
\usepackage[utf8]{inputenc}
\usepackage{amsmath,amssymb}

\usepackage{amsthm}
\usepackage{comment}

\usepackage{graphicx,subfig}
\usepackage[bbgreekl]{mathbbol}
\usepackage{bbm}

\usepackage{epsfig}

\usepackage{placeins}
\usepackage{float}
\usepackage{authblk}

\usepackage[ruled,vlined]{algorithm2e}
\usepackage{booktabs}

\def\ext{{\rm ext}}
\def\ex{{\rm ex}}
\def\bs{\boldsymbol}

\def\vp{{\varphi}}
\def\ve{{\varepsilon}}

\def\bi{{\bs i}}
\def\bj{{\bs j}}

\def\cB{\mathcal{B}}

\def\cF{\mathcal{F}}

\def\cT{\mathcal{T}}

\def\cW{\mathcal{W}}

\def\bfW{{\bf W}}
\def\bfZ{{\bf Z}}

\def\bRR{\mathbbm{R}}
\def\bZZ{\mathbbm{Z}}

\def\one{{\mathbbm{1}}}

\usepackage{ esint }

\def\pic{{\rm pic}}
\def\fbl{{\rm fbl}}

\def\rmd{\, {\rm d}} %for integrals
\def\Dt{{\Delta t}}
\def\Dx{{\Delta x}}

\def\dt{{\partial_{t}}}
\def\dx{{\partial_{x}}}

\def\dv{{\partial_{v}}}

\def\df{{\delta f}}

\def\beq{\begin{equation}}
\def\eeq{\end{equation}}

\providecommand{\abs}[1]{\lvert#1\rvert}

\providecommand{\norm}[1]{\lVert#1\rVert}

\DeclareMathOperator{\Div}{div}

\DeclareMathOperator{\remap}{\mathsf{Remap}}
\DeclareMathOperator{\field}{\mathsf{E}}

\begin{document}

\title{A $\df$ PIC method with Forward-Backward Lagrangian reconstructions}

\date{\today} % \\ $~$ \\ {\em Draft version -- please do not circulate}}

 \author[1]{Martin Campos Pinto}
 \author[2]{Merlin Pelz}
 \author[3]{Pierre-Henri Tournier}

\affil[1]{Max-Planck-Institut f\"ur Plasmaphysik, 
Boltzmannstraße 2,
D-85748 Garching b. München, Germany}
\affil[2]{Department of Mathematics, University of British Columbia, Vancouver, British Columbia, V6T 1Z2, Canada}
\affil[3]{Sorbonne Université, CNRS, Université de Paris, Laboratoire Jacques-Louis Lions (LJLL), F-75005 Paris, France}

\maketitle

\begin{abstract}
In this work we describe a $\df$ particle simulation method where the bulk density is periodically
remapped on a coarse spline grid using a Forward-Backward Lagrangian (FBL) approach.
This method is designed to handle plasma regimes where the densities strongly deviate 
from their initial state and may evolve into general profiles.
We describe the method in the case of an electrostatic particle-in-cell scheme and validate its qualitative 
properties using a classical two-stream instability subject to a uniform oscillating drive.
\end{abstract}

% \maketitle

\section{Introduction}

In order to reduce the statistical noise in numerical simulations of kinetic plasma 
problems, particle-in-cell (PIC) methods often follow a so-called $\df$ approach 
\cite{Kotschenreuther.1988.BAPS,Dimits_Lee_1993,Parker_Lee_1993} which consists 
of decomposing the transported density in two parts, a bulk density $f_0$ given 
by an analytical formula and a variation $\df$ represented with numerical particles. 
In Ref.~\cite{Aydemir_1994} this approach was revisited as a variance reduction method
in the scope of Monte Carlo algorithms, with $f_0$ playing the role of a control variate, and since 
then several techniques have been devised to improve the reduction of statistical error, 
in particular for gyrokinetic simulation models \cite{HTKKA_2002_pop,MHK_2004_pop,MBHSKK_2017,KHHMBBS_2019} 
and collisional models \cite{Chen.Parker.2007.PoP,VBVMJTBG_2010_pop,SWHK.2015.jcp}.

In many practical problems the plasma either remains close to 
an equilibrium state 
\cite{HTKKA_2002_pop, brochard_linear_2020},
or evolves as a small perturbation of some analytically known background
\cite{lanti_orb5_2020,biancalani_gyrokinetic_2021,rettino_gyrokinetic_2022}
which can be used as a bulk density.
In some cases however the plasma evolves in an unpredictable way and $f_0$ needs 
to be updated by a self-consistent algorithm to better follow the total density.

Typical examples are problems where full-$f$ simulations are needed.
In the physics of tokamak plasmas, one instance is the modeling of
$E \times B$ staircases \cite{dif-pradalier_finding_2015} which are long-lived patterns 
of quasiregular step-like profiles that develop spontaneously in turbulent plasmas. 
As they slowly move in the radial direction over large time scales, 
the separation assumption between an analytical background and fluctuations is no longer valid after 
some time and simulations need to involve full-$f$ methods such as the semi-Lagrangian scheme used in the 
GYSELA code \cite{grandgirard_5d_2016}.
Another example is the modelling of the tokamak edge region where the plasma density may
strongly deviate from local Maxwellian distributions, with large and intermittent fluctuations. 
This has motivated the development of various Eulerian full-$f$ schemes such as those presented in Ref.~\cite{dorf_progress_2020,mandell_electromagnetic_2020}, where
large plasma blob structures can be seen propagating from the core 
region towards the tokamak edge.

If one desires to model such problems with a $\df$ PIC method, it is thus necessary 
to allow for general updates of the bulk density over time.
An interesting approach in this direction was proposed in Ref.~\cite{Allfrey_Hatzky_2003}:
it consisted of projecting the particle 
density $\df$ on a coarse spline basis and add the resulting
smoothed distribution to the bulk density.

In this article we consider a variant of this approach where the bulk density is updated
using a semi-Lagrangian approach based on the Forward-Backward Lagrangian (FBL) method
\cite{Campos-Pinto.Charles.2018.jcm}. 

Inspired by Ref.~\cite{colombi_metric_2017}, the core of the FBL method 
is to compute backward trajectories on arbitrary nodes by local inversions of the 
particle trajectories: as these describe the forward transport flow in phase space and are 
naturally provided by the PIC code, their local inversion allows to perform semi-Lagrangian 
updates of a smooth density represented on a coarse grid. 
In particular the novelty of our approach is that it does not primarily rely 
on an accurate particle approximation of the density itself,
but rather on an accurate description of the particle trajectories.
As these are in general much less noisy than the phase space density, we believe
that this new paradigm can lead to efficient low-noise particle methods.

The outline is as follows.
In Section~\ref{sec:ansatz} we present our general ansatz for the discrete density,
which may be seen as a hybrid discretization between particle and semi-Lagrangian 
density representations. In Section~\ref{sec:transp} we recall the key steps of 
electrostatic particle-in-cell approximations, and in Section~\ref{sec:df-pic} we 
describe the $\df$ PIC method with FBL remappings of the bulk density.
The proposed method is summarized in Section~\ref{sec:over_df},
and in Section~\ref{sec:num} we present a series of numerical results involving
two-stream instability test cases to illustrate the enhanced denoising properties 
of our approach. 
We conclude in Section~\ref{sec:conclu} with a summary of the proposed method, 
a discussion on its novelty and two perspectives for future research.

\section{Approximation ansatz} 
\label{sec:ansatz}

Our ansatz for the general density at a discrete time $t^n$ is a sum of two terms,
\begin{equation} \label{ans}
  f^n := f^n_* + \df^n 
\end{equation}
(we shall often use $:=$ to highlight a definition) 
where the first one will be seen as the bulk density, a smooth approximation to the full solution,
and the second one as the fine scale variations.
Following the general principle of $\df$ methods we require $f^n_*$ to have a simple 
expression that is easy to evaluate at arbitrary positions in a general $d$-dimensional phase space, 
and we represent the variation $\df^n$ as an unstructured collection of numerical particles
with coordinates $z^n_k$ and weights $\delta w^n_k$,
\begin{equation} \label{dfn}
\df^n(z) := \sum_{k=1}^{N_p} \delta w^n_k \vp_\ve(z-z^n_k).
\end{equation}
Here $\vp_\ve$ is a smooth shape function of scale $\ve$ and $z = (z_1, \dots, z_d)$ is a phase-space coordinate.
In typical problems where the transported density slightly deviates from an initial
profile, the bulk density is often set to this initial value, $f^n_* = f^0$ or to some analytical 
equilibrium \cite{HTKKA_2002_pop, brochard_linear_2020}.
In this note we investigate an alternate approach where 
the bulk density is represented as an arbitrary collection of 
B-splines on a coarse grid with mesh-size $h_* > \ve$,
in the spirit of Ref.~\cite{Allfrey_Hatzky_2003}.
This yields an expression that is formally similar to that of $\df^n$,
\begin{equation} \label{f*n}
f^n_*(z) = \sum_{\bj \in \bZZ^d} w^n_{*,\bj} \vp_{h_*}(z-\bj h_*)
\end{equation}
where $\vp_*(\cdot - \bj h_*)$ is now the coarse B-spline shape centered on a 
general $d$-dimensional grid node $\bj h_*$ 
($\bZZ$ is the set of integers) and $w^n_{*,\bj}$ is its weight.
In practice, only a finite number $N_*$ of such nodes is used, and 
as outlined in the introduction we will periodically update the coefficients of 
this spline bulk density using a Forward-Backward Lagrangian (FBL) reconstructions, 
which involves a relatively small set of passive markers designed to track the characteristic flow in phase space.

\subsection{Main numerical parameters and limit regimes}
\label{sec:params}

The main numerical parameters are as follows.

\begin{itemize}
    \item $N_r$ is the remapping period, i.e., the number of time steps between two 
    updates of the bulk density. 
    The limit value $N_r = \infty$ corresponds to a frozen bulk density, namely $f^n_* = f^0_*$ for every time step $n$.
    \item $N_*$ is the number of coarse splines used to represent the bulk density in the computational domain $\Omega \subset \bRR^d$,
    it is on the order of ${\rm Vol}(\Omega)(h_*)^{-d}$.
    The limit value of $N_* = 0$ (an empty grid) corresponds to a ``full-$f$'' particle approximation. 
    \item $N_p$ is the number of numerical particles describing the fine scale structures.
    The limit value of $N_p = 0$ corresponds to a semi-Lagrangian ansatz \cite{Sonnendrucker.Roche.Bertrand.Ghizzo.1999.jcp} 
    where the full density is represented on a structured grid and can be updated in time with an ad-hoc scheme such as the FBL method 
    described in Ref.~\cite{Campos-Pinto.Charles.2018.jcm}.
\end{itemize}

\subsection{Weighted collections of spline shape functions}

For simplicity, we consider spline shape functions for both the bulk density and the fine scale variations.
Specifically, we set 
$$
\vp_\ve(z) := \Big(\frac{1}{\ve}\Big)^d \vp\left(\frac{z}{\ve}\right), \qquad z \in \bRR^d
$$
with a reference shape function $\vp$ defined as a centered cardinal 
B-spline of degree $p$,
\begin{equation*}
\vp(z) := \prod_{i=1}^d B_p(z_i) \quad \text{ with support } ~~ 
\Big[-\frac{p+1}{2},\frac{p+1}{2}\Big]^d
\end{equation*}
involving standard univariate B-splines 
defined recursively by
\begin{equation*}
B_0(x) := \one_{\left[-\tfrac 12, \tfrac 12\right]}(x)
~~ \text{ and } ~~
B_p(x)
:= \int_{x-\frac 12}^{x+\tfrac 12} B_{p-1}
~~ \text{ for } ~ p \ge 1.
\end{equation*}
In the sequel, it will be convenient to denote an arbitrary collection of weighted splines as
\begin{equation*} 
    \Phi_\ve[\bfW,\bfZ](z) := \sum_{k = 1}^N w_k \varphi_\ve(z-z_k)
\end{equation*}
where $\bfW = (w_k)_{k= 1 \dots N}$, $\bfZ = (z_k)_{k= 1 \dots N}$.
With this convention, the two components of our general ansatz \eqref{ans} read
\begin{equation} \label{ans-Phi}
    f^n_* = \Phi_{h_*}[\bfW^n_*,\bfZ_*] 
    \qquad \text{ and } \qquad
    \df^n = \Phi_\ve[\delta\bfW^n,\bfZ^n]
\end{equation} 
where $\bfZ_* := (\bj h_*)_{\bj \in \bZZ^d}$ are the nodes of the spline grid.

\section{Particle approximations to transport equations} 
\label{sec:transp}

Our method may be described for general non-linear transport problems of the form 
\begin{equation} \label{transport}
    \dt f(t,z) + U[f]\cdot \nabla_z f(t,z) = 0
\end{equation}
where $z \in \bRR^d$ is the phase-space variable and $U[f]$ the generalized velocity
field associated to the solution $f$.

\subsection{Characteristic flows}
The characteristic trajectories associated to Eq.~\eqref{transport}
are the curves $Z(t) = Z(t;s,z) \in \bRR^d$, solution to the ODEs
$$
\frac{\rmd}{\rmd t} Z(t) = U[f](t,Z(t)), \qquad Z(s) = z
$$
for arbitrary $s,t \in [0,T]$ and $z \in \bRR^d$, see e.g. Ref.~\cite{Raviart.1985.lnm}.
The (forward) characteristic flow between two times $t^{n} = n\Dt$ and $t^{n+1} = (n+1)\Dt$
is then defined as
\begin{equation}\label{Fex}
    \cF_\ex^{n,n+1}(z) := Z(t^{n+1};t^{n},z)
\end{equation}
and the inverse mapping $\cB_\ex^{n,n+1} := \big(\cF_\ex^{n,n+1}\big)^{-1}$ is the backward flow 
\begin{equation}\label{Bex}
\cB_\ex^{n,n+1}(z) = Z(t^n;t^{n+1},z).
\end{equation}
Using the backward flow we can write the solution to Eq.~\eqref{transport} over the time interval
$[t^n,t^{n+1}]$  as
\begin{equation} \label{TfB}
    f(t^{n+1},z) = f(t^{n},\cB^{n,n+1}_\ex(z)).
\end{equation}
Here we may restrict ourselves to divergence-free velocity fields: $\Div_z U[f] = 0$. 
The characteristic flows are then measure-preserving and the transport is conservative.

\subsection{1D1V Vlasov-Poisson equation}
A simple example is the periodic Vlasov-Poisson equation 
in a two dimensional phase-space, i.e., $z = (x,v)$ with a 
periodic space coordinate $x \in [0,L]$ and velocity $v \in \bRR$
\begin{equation} \label{VP}
\dt f(t,x,v) + v \dx f(t,x,v) - E(t,x) \dv f(t,x,v) = 0
\end{equation}
for $t \ge 0$, $(x,v) \in [0,L] \times \bRR$,
with a normalized initial density $\int_0^L\int_\bRR f^0(x,v) \rmd v\rmd x  = 1$ and a periodic electric field $E = E[f]$ defined by
\begin{equation} \label{Ef}
    \left\{
    \begin{aligned}
     E(t,x) &= - \dx \phi(t,x)
    \\
     -\Delta \phi(t,x) &= \rho(t,x) = \frac 1L - \int_\bRR f(t,x,v) \rmd v.
    \end{aligned}
    \right.
\end{equation}
This corresponds to an electrostatic, normalized ($\ve_0 = q_e = m = 1$) periodic electron plasma in 1D,
with constant neutralizing background ion density, so that $\int_0^L \rho \rmd x  = 0$. Here the generalized velocity field is
$$
U[f](t,z) = \big( v, -E(t,x) \big) \qquad  \text{ with } ~  z = (x,v).
$$
Due to the non-linear nature of this transport equation, the characteristic flow
has no explicit expression.

\subsection{Full-$f$ particle approximation}

Particle approximations represent the transported density $f^n(z) \approx f(t^n,z)$ 
as a sum of numerical particles of the form 
\begin{equation} \label{fn}
    f^n(z) = \Phi_\ve[\bfW^n,\bfZ^n](z)    
\end{equation}
with weights initially set to
\begin{equation} \label{w}
    w^0_k := \frac{f^0(z^0_k)}{N_p g^0(z^0_k)}, \qquad k = 1, \dots, N_p,
\end{equation}
where $g^0$ is the sampling distribution of the initial markers $\bfZ^0 = (z^0_k)_{k=1,\dots, N_p}$, 
see e.g. Ref.~\cite{SWHK.2015.jcp}.
As the problem is conservative the weights are kept constant in time, $\bfW^{n+1} = \bfW^n$,
and the markers are pushed forward
\begin{equation} \label{push}
\bfZ^{n+1} = \cF^n(\bfZ^n)
\end{equation}    
using some approximation to the forward flow in Eq.~\eqref{Fex} which takes as parameter 
the numerical solution at time $t^n$,
$$
\cF^n(\bfZ) = \cF_{\Dt}[\bfW^n,\bfZ^n](\bfZ).
$$

\subsection{Electrostatic full-$f$ leap-frog flow}

For the Vlasov-Poisson equation \eqref{VP}, a standard numerical flow
is given by a leap-frog (Strang splitting) scheme,
\begin{equation} \label{LF-F}
  \cF^{\rm lf, \pic}_{\Dt,\Dx}[\bfW^n,\bfZ^n](\bfZ) = \cF^{\rm x}_{\frac \Dt 2}
      \circ \cF^{\rm v, \pic}_{\Dt,\Dx}[\bfW^{n+\frac 12},\bfZ^{n+\frac 12}] \circ \cF^{\rm x}_{\frac \Dt 2}(\bfZ)
\end{equation}
(where $\circ$ is the usual function composition operator)
associated with a grid with $M$ points and step-size
$\Dx = L/M$.
The first split flow reads
\begin{equation} \label{LF-x}
\cF^{\rm x}_{\frac \Dt 2}(x,v) := \big(x + \tfrac {\Dt}{2} v, v\big)
\end{equation}
and the second one takes the form
\begin{equation} \label{LF-v}
\cF^{\rm v, \pic}_{\Dt,\Dx}[\bfW^{n+\frac 12},\bfZ^{n+\frac 12}](x,v) := \big(x, v - \Dt E^{n+\frac 12}(x) \big).
\end{equation}
Here we have set
\begin{equation} \label{WZ_half}
\bfW^{n+\frac 12} := \bfW^{n},
\qquad\qquad
\bfZ^{n+\frac 12} := \cF^{\rm x}_{\frac \Dt 2}(\bfZ^n)
\end{equation}
and the electric field is computed from the particles with a discrete Poisson solver
such as the one described just below,
that takes as input the particle weights and positions,
\begin{equation} \label{LF-E}
E^{n+\frac 12}(x) := \field^{\pic}_{\Dx}[\bfW^{n+\frac 12},\bfZ^{n+\frac 12}](x).
\end{equation}

\subsection{Electrostatic full-$f$ PIC field solver}
\label{sec:LF-pic}

To clarify how the particles are coupled with the field in Eq.~\eqref{LF-E},
we recall the precise form of a basic Finite Difference (FD) Poisson solver
for a general collection of weighted particles, i.e.,
\begin{equation}\label{Esolve_pic}
    \Phi_\ve[\bfW, \bfZ] \mapsto \field^{\pic}_{\Dx,\ve}[\bfW, \bfZ].
\end{equation}
We note that for periodic systems spectral solvers involving fast Fourier transforms are often 
used, see e.g. Ref.~\cite[Sec.~8]{Birdsall_Langdon_1991}. 
In both cases the charge deposition algorithm consists of computing grid values for the charge density, 
\begin{equation} \label{rhopic}
\rho_i = \frac 1L - \int_\bRR \Phi_\ve[\bfW, \bfZ](i\Dx,v) \rmd v = \frac 1L - \sum_{k=1}^{N_p} w_k \vp_\ve(i \Dx - x_k)
\end{equation}
for $i = 1, \dots M$.
A discrete Poisson equation is then solved. With a FD scheme this reads
$$
E_i = -\frac{\phi_{i+1}-\phi_{i-1}}{2\Dx},
\qquad
-\frac{\phi_{i+1}-2\phi_{i}+\phi_{i-1}}{\Dx^2} = \rho_i,
$$
for $i = 1, \dots M$.
The electric field \eqref{Esolve_pic} is then defined using the same spline shape 
functions as the particles but a different normalization, and scaled with the FD grid:
\begin{equation*} 
    \field^{\pic}_{\Dx,\ve}(x) := \sum_{i=1}^{M} E_i \tilde \vp_\Dx (x - i\Dx) \qquad \text{where} \quad
    \tilde \vp_\Dx(x) = \vp\Big(\frac{x}{\Dx}\Big).
\end{equation*}
In the usual case where $\ve = \Dx$, we write $\field^{\pic}_{\Dx} = \field^{\pic}_{\Dx,\Dx}$.

\section{A $\df$ PIC scheme with remappings}
\label{sec:df-pic}

We now describe the main steps of a $\df$ method involving
a decomposition of the form Eq.~\eqref{ans}. 
Denoting by $N_r$ the remapping period, we first observe that 
the bulk part is frozen between two remapping steps. This yields in particular 
\begin{equation} \label{frozen}
  f^n_* = f^m_*
  \qquad \text{ where } \qquad
  m = m(n) := \left\lfloor \frac{n}{N_r}\right\rfloor N_r.
\end{equation}
The $\df$ part is then evolved as follows. 

\subsection{$\df$ particle approximation: basic steps}

A typical $\df$ time step consists of 
(i) pushing the markers forward to follow the characteristic
curves as in Eq.~\eqref{push},
\begin{equation} \label{push_df}
\bfZ^{n+1} = \cF^n(\bfZ^n)
\end{equation}    
and (ii) updating the weights. 
Indeed, since 
the numerical particles now represent a variation $\df^n = f^n - f^n_*$
with frozen $f^n_*$ on the time step, 
their weights no longer carry a conserved density and they must be evolved in time.
In the {\em direct} $\df$ method
(see e.g. Ref.~\cite[Sec.~3]{SWHK.2015.jcp}) they are set as
\begin{equation} \label{dw}
  \delta w^n_k := \frac{f^n_k-f^n_*(z^n_k)}{N_p g^n_k}
  \approx \frac{\df(t^n,z^n_k)}{N_p g(t^n,z^n_k)}, \qquad k = 1, \dots, N_p
\end{equation}
where
\begin{equation} \label{fgk}
f^n_k \approx f(t^n,z^n_k)
\qquad \text{ and } \qquad
g^n_k \approx g(t^n,z^n_k)
\end{equation}
correspond to the transported density and the markers density probability, respectively, evaluated at the particle positions.
Since both values are constant along exact trajectories, a reasonable choice is
\begin{equation} \label{fgk0}
    f^n_k := f^0(z^0_k)
    \qquad \text{ and } \qquad
    g^n_k := g^0(z^0_k).
\end{equation} 
Up to these constant values 
we see that the $\df$ weights only depend on the bulk density,
which is characterized by the weights $\bfW^n_* = \bfW^m_*$, at the 
marker positions $\bfZ^n$. Hence we may summarize the weighting scheme in Eq.~\eqref{dw}--\eqref{fgk0} as 
\begin{equation} \label{dW}
  \delta \bfW^n := \delta\cW(\bfW^m_*,\bfZ^n).
\end{equation}

\subsection{Electrostatic $\df$-PIC flow}
\label{sec:LF-df}

In the case of a $\df$ ansatz, the leap-frog flow reads
\begin{equation} \label{LF-df-F}
  \cF^{{\rm lf},\df}_{\Dt,\Dx}[\bfW^m_*,\bfZ^n] = \cF^{\rm x}_{\frac \Dt 2}
      \circ \cF^{{\rm v},\df}_{\Dt,\Dx}[\bfW^m_*,\bfZ^{n+\frac 12}] \circ \cF^{\rm x}_{\frac \Dt 2}
\end{equation}
with split flows defined similarly as in Eq.~\eqref{LF-x}--\eqref{LF-v}
and a field solver adapted to the ansatz from Eq.~\eqref{ans}, namely
\begin{equation} \label{LF-df-v}
    \cF^{\rm v, \df}_{\Dt,\Dx}[\bfW^m_*,\bfZ^{n+\frac 12}](x,v) := \big(x, v - \Dt E^{n+\frac 12}(x) \big)
\end{equation}
with 
\begin{equation} \label{E_*df}
    E^{n+\frac 12} = E^m_{*} + E_{\df}^{n+\frac 12} = \field_{\Dx,h_*}[\bfW^m_*,\bfZ^{n+\frac 12}].
\end{equation}
Since both parts of the density are formally written as weighted collections of spline shapes,
we may use a standard PIC solver for each parts.
The field induced by the bulk density $f^m_*$ is
\begin{equation} \label{E*}
E^m_{*} := \field^{\pic}_{\Dx,h_*}[\bfW^m_*,\bfZ_*]
\end{equation}
and the field corresponding to the variation
$\df^{n+\frac 12} = \Phi_\ve[\delta\bfW^{n+\frac 12},\bfZ^{n+\frac 12}]$,
with scaling $\ve = \Dx$, reads
\begin{equation} \label{dE}
\begin{aligned}
&E_{\df}^{n+\frac 12} = \field^{\pic}_{\Dx}[\delta \bfW^{n+\frac 12},\bfZ^{n+\frac 12}]
\\ \noalign{\medskip}
\text{with }& \qquad
\left\{
\begin{aligned}
  \delta \bfW^{n+\frac 12} &:= \delta\cW(\bfW^m_*,\bfZ^{n+\frac 12})
  \\
  \bfZ^{n+\frac 12} &:= \cF^{\rm x}_{\frac \Dt 2}(\bfZ^n)
\end{aligned}
\right.
\end{aligned}
\end{equation}
see \eqref{WZ_half}, \eqref{dW}, \eqref{dw}.
The resulting field solver corresponding to \eqref{E_*df} may then be written as
\begin{equation*} 
\begin{aligned}
\field_{\Dx,h_*}[\bfW^m_*,\bfZ^{n+\frac 12}] 
:= & \field^{\pic}_{\Dx,h_*}[\bfW^m_*,\bfZ_*] \\ 
   & + \field^{\pic}_{\Dx}[\delta\cW(\bfW^m_*,\bfZ^{n+\frac 12}),\bfZ^{n+\frac 12}].
\end{aligned}
\end{equation*}

Since the field $E^m_*$ is constant between two remappings, 
regular time steps only involve a deposition of the $\df$ particles to
update the $E_{\df}$ part.
Note that even if $\abs{\df^n} \ll \abs{f^m_*}$, there is no reason
why $E_{\df}$ should be small compared to $E^m_*$. In particular,
we have $E^m_* \approx 0$ when $f^m_*$ is close to an equilibrium density.

\subsection{FBL remappings with auxiliary markers}

For the remappings, namely the updates of the bulk density $f^{m}_* \mapsto f^{n}_*$ with $n = m + N_r$, 
we resort to Forward-Backward Lagrangian (FBL) reconstructions \cite{Campos-Pinto.Charles.2018.jcm}.
As mentioned above this involves a collection of auxiliary markers that have been reset on the Cartesian grid
at the previous remapping step,
\begin{equation} \label{remaptZh}
    \tilde \bfZ^m := \bfZ_*,
    \qquad {i.e.},
    \qquad \tilde z^m_\bj := \bj h_* \quad \forall \bj \in \bZZ^d,
\end{equation}
and pushed forward similarly as the classical ones in Eq.~\eqref{push_df},
\begin{equation} \label{tZm}
    \tilde \bfZ^{n} = \cF^{n-1} \big( \cdots \cF^{m}(\bfZ_*) \cdots \big)    
\end{equation}
in order to track the forward characteristic flow. The FBL method then 
performs a semi-Lagrangian step 
\begin{equation} \label{f*SL}
  f^n_* := A_* \cT_{\rm fbl}[\tilde \bfZ^{n}] f^{m}_* 
\end{equation}
where $A_*$ is a spline interpolation or quasi-interpolation operator on the $h_*$ grid, and
$
\cT_{\rm fbl}[\tilde \bfZ^{m}] : f \mapsto f \circ \cB^{m,n}_{\fbl}
$ 
is a transport operator approximating the exact one in Eq.~\eqref{TfB}, based on an FBL backward flow
\begin{equation} \label{B_fbl}
    \cB^{m,n}_\fbl = \cB_\fbl[\tilde \bfZ^{n}] \approx \cB^{m,n}_\ex.    
\end{equation}
As a result we obtain a numerical scheme to update the coefficients of the bulk density, which we summarize as 
\begin{equation} \label{remapW*} 
  \bfW^n_* := \remap[\tilde \bfZ^n](\bfW^{m}_*)
\end{equation}
where $n$ is a multiple of $N_r$ and $m = n-N_r$.

In the next subsection we give some details about the FBL algorithm.

\subsection{FBL density reconstructions}
\label{sec:fbl}

The FBL scheme in Eq.~\eqref{f*SL} performs a semi-Lagrangian approximation with a backward flow 
derived
from the position of the auxiliary markers $\tilde \bfZ^{m}$ in Eq.~\eqref{tZm}, as follows:
\begin{itemize}
    \item[(i)] For any index $\bj \in \bZZ^d$, let $\tilde z^m_\bi$ with $\bi = \bi(\bj)$  
    be an FBL marker close to the node $\bj h_*$ 
    \item[(ii)]
    Compute the quadratic local backward flow associated with $\tilde z^n_\bi$, namely
    \begin{equation} \label{BwdFlow}
         B^{m,n}_\bi(z) = z^*_\bi + D^n_\bi(z - \tilde z^n_\bi) + \tfrac 12 (z - \tilde z^n_\bi)^T Q^n_\bi(z - \tilde z^n_\bi)
    \end{equation}
    where $z^*_\bi = \bi h_*$, and where $D^n_\bi$, $Q^n_\bi$ correspond respectively 
    to the Jacobian matrix and Hessian tensor of the 
    backward flow at $z= \tilde z^n_\bi$, computed from the positions 
    of the logical neighbors of $\tilde z^n_\bi$ 
    (for more details see Ref.~\cite{Campos-Pinto.Charles.2018.jcm})
    \item[(iii)]
    Reconstruct the transported density at node $z^*_\bj = \bj h_*$ using this approximated flow,
    \begin{equation} \label{BSL_f*}
        f^n_{*,\bj} := f^{m}_*(B^{m,n}_\bi(z^*_\bj))
    \end{equation}
    \item[(iv)]
    Compute the new weights $\bfW^n_*$ using a spline approximation operator $A^*$.
\end{itemize}

For the operator $A_*$,
we can use a standard interpolation scheme, or even a local quasi-interpolation scheme,
see, e.g., Ref.~\cite{Chui.Diamond.1990.numat}. 
Both compute high-order B-spline approximations, but the latter only involves local pointwise evaluations
of the target function. The resulting approximation takes the form
\begin{equation}\label{A*}
\begin{aligned}
&A_* g(z) := \sum_{\bj \in \bZZ^d} w_\bj(g) \vp_{h_*}(z - \bj h_*)
\\
&\quad  \text { with } \quad
w_{\bj}(g) := h_*^d \sum_{\norm{\bi}_\infty \le s_p} a_{\bi} \, g((\bj+\bi)h_*),
\end{aligned}
\end{equation}
with coefficients $a_{\bi} := a_{i_1} \cdots a_{i_d}$
on a local stencil of size $s_p$, as
given by the symmetric method ($a_i = a_{-i}$) in Ref.~\cite[Sec.~6]{Chui.Diamond.1990.numat}.

\section{Overview of the $\df$ particle scheme with FBL remappings}
\label{sec:over_df}

Gathering the above steps, we may write the solution at time $t^n$
as
\begin{equation} \label{f*df}
  f^n = f^m_* + \df^n
  \qquad \text{ with } \qquad
  \left\{ \begin{aligned}
  & f^m_* = \Phi_{h_*}[\bfW^m_*,\bfZ_*] 
  \\
  & \df^n = \Phi_\ve[\delta\bfW^n,\bfZ^n].
\end{aligned}
\right.
\end{equation}
Here we remind that $\bfZ_* = (\bj h_*)_{\bj \in \bZZ^d}$ are the coarse grid nodes
and $m = m(n)$ is the last remapping step preceding $n$, see Eq.~\eqref{frozen}.
The proposed scheme reads then as follows.

\subsection{Initialization}

\begin{itemize}

    \item Compute the weights $\bfW^0_*$ of the initial bulk density $f^0_* = \Phi_{h_*}[\bfW^0_*,\bfZ_*]$ with
    $$
    f^0_* := A_* f^0
    $$
    using a spline approximation scheme such as the one in Eq.~\eqref{A*},

    \item Draw the $\df$ markers according to some initial sampling distribution
    $$
    \bfZ^0 \sim g^0
    $$

    \item Initialize the auxiliary FBL markers on the $h_*$ grid \eqref{remaptZh},
    $$
    \tilde \bfZ^0 := \bfZ_* ~.
    $$

\end{itemize}

\subsection{Time loop} 
For $n = 0, \cdots, N_t-1$,
let 
$
m := 
\big\lfloor \frac{n}{N_r}
\big\rfloor N_r
$
denote the previous remapping step as in Eq.~\eqref{frozen}, and do:

\begin{itemize}
    \item If $n = m > 0$, {\bf remap}:

    \begin{itemize}
        \item[$\triangleright$]
        Update the bulk density weights (coefficients) 
        with the FBL  method 
        described in Section~\ref{sec:fbl},
        \begin{equation}\label{remap}
            \bfW^m_* := \remap[\tilde \bfZ^m](\bfW^{\hat m}_*)            
        \end{equation}
        \item[$\triangleright$] Reset the FBL markers with \eqref{remaptZh}, $\tilde \bfZ^m := \bfZ_*$ 
    \end{itemize}

    \item {\bf Push} the markers forward:
    \begin{equation} \label{push_Z_tZ}
    \left\{
    \begin{aligned}
        &\bfZ^{n+1} := \cF^n(\bfZ^n)
        \\
        &\tilde \bfZ^{n+1} := \cF^n(\tilde \bfZ^n)
    \end{aligned}
    \right.
    \end{equation}
    using some discrete flow 
    $
    \cF^n = \cF^{\df}_{\Dt}[\bfW^m_*,\bfZ^n]
    $
    such as the leap-frog scheme in Eq.~\eqref{LF-df-F},
    which involves deposition and field solver steps
    as detailed in Section~\ref{sec:LF-df} and \ref{sec:LF-pic}.
\end{itemize}

\begin{figure*}
  \includegraphics{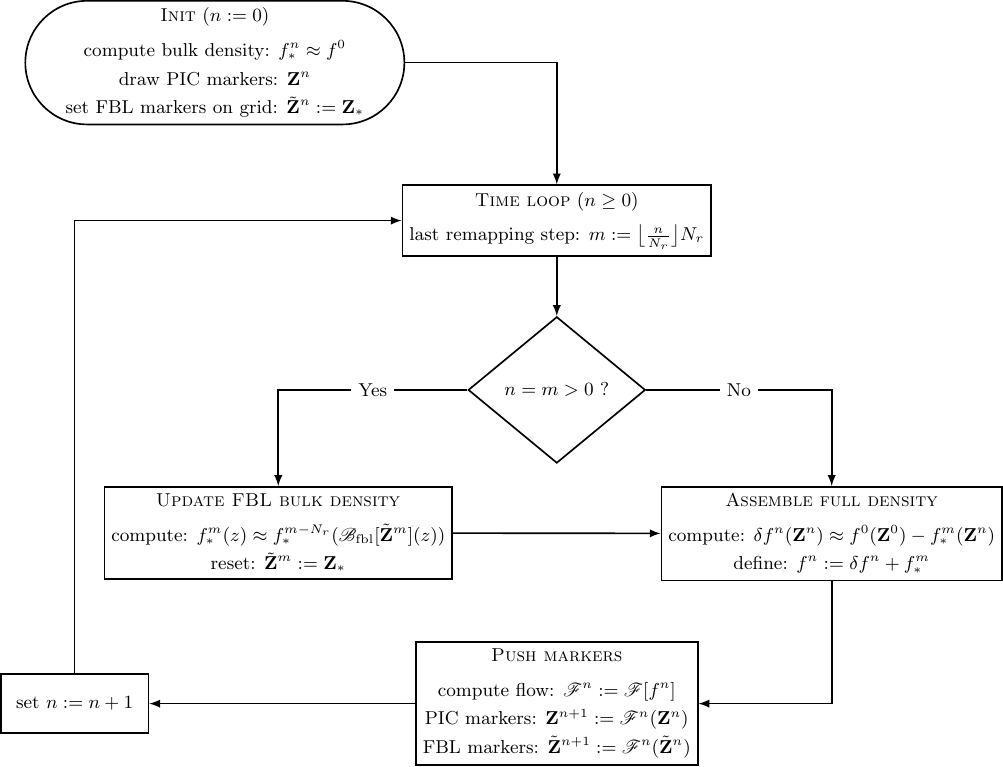}
  \caption{
  \label{fig:FBL_df_scheme}    
  Flowchart summarizing the main steps of the FBL-$\df$ scheme.
  }    
\end{figure*}

\subsection{FBL variant}
\label{sec:FBLvar}

A semi-Lagrangian FBL variant of the above scheme can be obtained by
(i) setting $N_p = 0$ and (ii) treating the FBL markers $\tilde \bfZ^n$ as ``active'' particles
in the field solver.

Specifically, this amounts to discarding the $\df$ part ($\bfZ^n = \emptyset$)
so that the full density is represented on the structured spline grid on the successive remapping steps, 
and to defining the electric field in Eq.~\eqref{LF-df-F}--\eqref{LF-df-v} by
\begin{equation} \label{E_fbl}
    E^{n+\frac 12} = \field^{\pic}_{\Dx}[\bfW^{n+\frac 12}_*,\tilde \bfZ^{n+\frac 12}]
\end{equation}
as one would do with a standard full-$f$ PIC scheme, see \eqref{LF-v}--\eqref{LF-E}, using the weighted FBL markers
as standard particles. Note that here the weights are those set at the last remapping step \eqref{remap}, i.e., 
$\bfW^{n+\frac 12}_* = \bfW^m_*$ with $m$ as in \eqref{frozen}.

\section{Numerical results}
\label{sec:num}

In this section we assess the proposed method on the periodic 1D1V Vlasov-Poisson system 
\eqref{VP}--\eqref{Ef},
where we compare it with other PIC and semi-Lagrangian schemes.

\subsection{Numerical parameters for the tested methods} 

For the purpose of comparison, we use different values for the main parameters listed in Section~\ref{sec:params}, 
including some limiting values which allows to emulate full-$f$ and $\df$ PIC schemes, as well as a semi-Lagrangian
FBL scheme. All the methods use cubic splines for the particle shape functions.

As the solutions are typically supported in a region occupying roughly one half
of the computational phase-space domain, for each grid resolution $N_x \times N_v$ we 
indicate as $N^{\rm eff}_* \approx N_*/2$, with $N_* = N_x N_y$, 
the approximate number of active spline coefficients on the grid, corresponding to the ``effective'' grid resolution.
The resulting methods are summarized in Table \ref{tab:methods}.

\begin{table}[!h]
  \caption{Numerical parameters used for the different methods, as described in the text.
  }
  \label{tab:methods}
  \begin{center}
  \begin{tabular}{|c|c|c|c|c|c|}
    \hline
    method & $N_p$ & grid (FBL) & 
        $N^{\rm eff}_*$ & $N_r$ & FBL markers
    \\
    \hline
    \hline
    full-$f$ PIC, coarse & 20,000 & $0 \times 0$ 
        & 0 & $\infty$ & $\emptyset$
    \\
    \hline
    full-$f$ PIC, fine & 320,000 & $0 \times 0$ 
        & 0 & $\infty$ & $\emptyset$
    \\
    \hline
    $\df$ PIC & 18,752 & $50 \times 50$
        & 1,250 & $\infty$ & frozen
    \\
    \hline
    FBL, coarse & 0 & $200 \times 200$ 
        &  20,000 & $20$  & active
    \\
    \hline
    FBL, fine & 0 & $800 \times 800$ 
        &  320,000 & $20$  & active
    \\
    \hline
    FBL-$\df$ PIC & 18,752 & $50 \times 50$ 
        &  1,250 & $20$ & passive
    \\ 
    \hline
  \end{tabular}
\end{center}
\end{table}
Here the active/passive/frozen status of FBL markers $\tilde \bfZ^n$ refers to their role in the 
scheme and in particular in the field solver that is involved in the push step in Eq.~\eqref{push_Z_tZ}:
{\em active} markers (used in the FBL variant) carry a charge which is deposited as explained in Section~\ref{sec:FBLvar}
while {\em passive} markers (used in the proposed FBL-$\df$ scheme) do not, as described in Section~\ref{sec:LF-df}. 
Finally {\em frozen} markers are passive, and never pushed: in the absence of remappings only their initial position on the grid 
is used to represent the bulk density $f^n_* = f^0_*$.

We point out that in this study we do not attempt to assess the computational cost of the various methods, as the problems 
are academic and our implementation is far from being optimal. In particular one should keep in mind that 
full-$f$ and basic $\df$ methods, even if noisier, may be quicker to run and easier to parallelise than their FBL
counterparts. Efficient implementations of the latter are left for future works.

\subsection{Test cases}

Our numerical experiments are based on a 1D1V two-stream instability test case
where the initial distribution is of the form
\begin{equation} \label{f0_tsi}
   f^0(x,v) = ( 1 + \alpha \cos(k x) ) \frac{v^2  }{\sqrt{2\pi}} e^{-\frac{v^2}{2}}
\end{equation}
with a wave number of $k = \frac 12$ and a perturbation amplitude of $\alpha = 0.01$.
This solution is known to develop thin filaments in the phase space that are difficult to resolve numerically,
see, e.g., Refs.~\cite{Cheng_Knorr_1976,Rossmanith.Seal.2011.jcp}. Here we will simulate two-stream instabilities
on the time interval $[0,100]$. We shall consider two versions of the problem: 
\begin{itemize}
    \item[(i)] the basic Two-Stream Instability (TSI) test case where the solution is evolved with the self-consistent
    Vlasov-Poisson system  \eqref{VP}--\eqref{Ef}
    \item[(ii)] a Driven Two-Stream Instability (DTSI) test case where an oscillating external field 
    \begin{equation} \label{drive}
        E_\ext(t,x) = \frac {1}{10} \cos\Big(\frac{\pi t}{50}\Big) 
    \end{equation}
    is added to the self-consistent field from Eq.~\eqref{Ef}.    
    The effect of this external drive is to add a periodic perturbation to the TSI trajectories, uniform in $x$ and $v$, 
    $$
    (x,v) \mapsto \left(x+\frac{250}{\pi^2}\Big(1-\cos\Big(\frac{\pi t}{50}\Big)\Big), y+\frac{5}{\pi}\sin\Big(\frac{\pi t}{50}\Big)\right),
    $$ 
    so that the solutions to both test cases coincide for $t = 100$.
\end{itemize}
The computational phase-space domain is set to 
$$\Omega = \big[0,L\big] \times [-v_{\rm max},v_{\rm max}]
$$
with $L = \frac{2\pi}{k}$ and $v_{\rm max} = 10$. 
Note that this value is significantly larger than what is usually seen in the literature,
in order to correctly represent the variations in velocity induced by the oscillating drive. 
Finally the initial sampling distribution for the particles is taken as
$g^0 = (\int f^0(z)\rmd z)^{-1}f^0$ for the full-$f$ PIC scheme, and as 
\begin{equation} \label{g_gauss}
    g^0(x,v) = \frac{1}{L} \frac{1}{\sigma\sqrt{2\pi}}e^{-\frac{v^2}{2\sigma^2}}
\end{equation}
with $\sigma = 2$, for the $\df$ and FBL-$\df$ PIC methods.

\subsection{Phase-space density plots}

\begin{figure*} [!htbp]
\centering
\includegraphics[width=\textwidth]{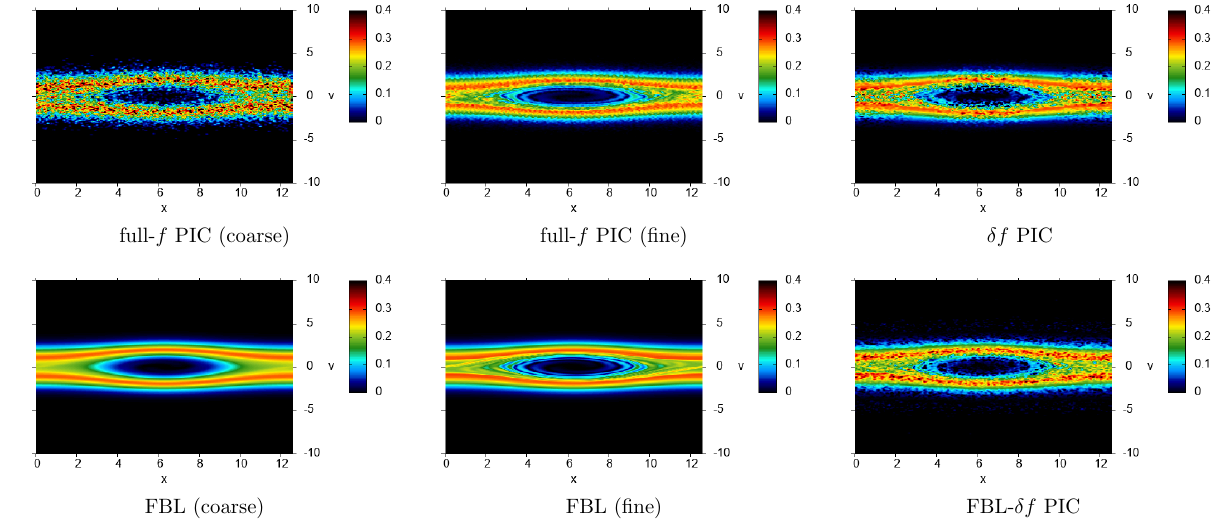}
\caption{ \label{fig:TSI_f}
Phase-space densities $f^n$ obtained for the TSI test case at $t^n = 100$ 
with the methods listed in Table~\ref{tab:methods}. Here the basic $\df$ and FBL-$\df$ methods use about 18,750 standard 
particles and 1,250 FBL markers, which is comparable to the 20,000 particles (or active markers) used in 
the coarse PIC and FBL methods. Fine PIC and FBL simulations use 16 as many particles.
}  
\end{figure*}

\begin{figure*} [!htbp]
\includegraphics[width=\textwidth]{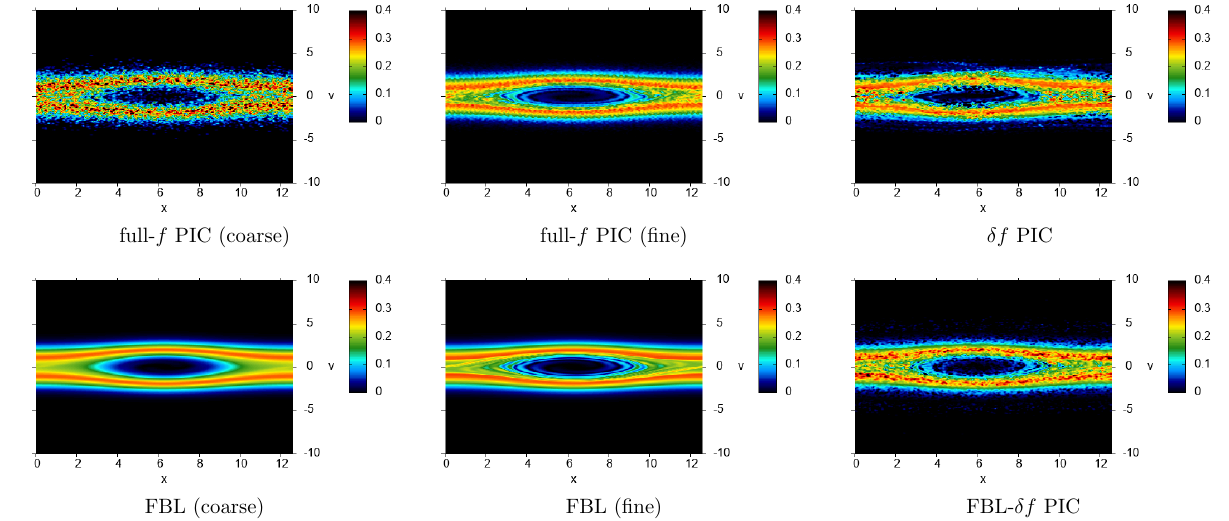}
  \caption{
   Densities obtained for the DTSI test case at $t^n = 100$,
   where the exact solution coincides with the TSI one. 
   As in Fig.~\ref{fig:TSI_f}, the different methods are those listed in Table~\ref{tab:methods}.
  }
\label{fig:DTSI_f}
\end{figure*}

\begin{figure*} [!htbp]
\includegraphics[width=\textwidth]{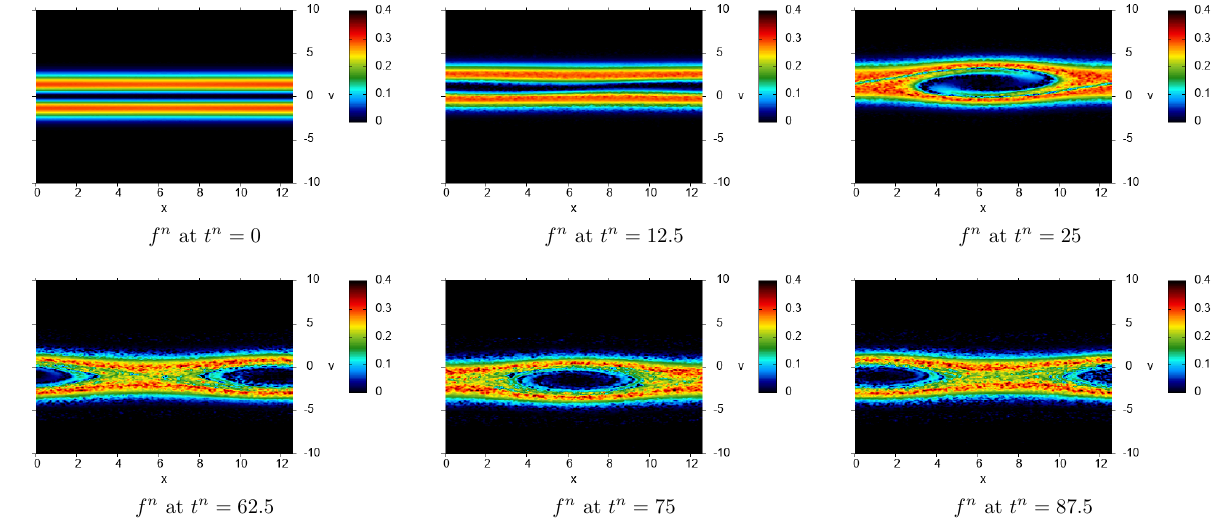}
  \caption{
  Snapshots of the density $f^n$ computed by the FBL-$\df$ PIC scheme at various times $t^n$, for the DTSI test case.
  At time $t^n = 100$, the corresponding density is shown in the lower-right plot of Fig.~\ref{fig:DTSI_f}.
  }
  \label{fig:DTSI_FBL_df_evol_f}
\end{figure*}

\begin{figure*} [!htbp]
  \includegraphics[width=\textwidth]{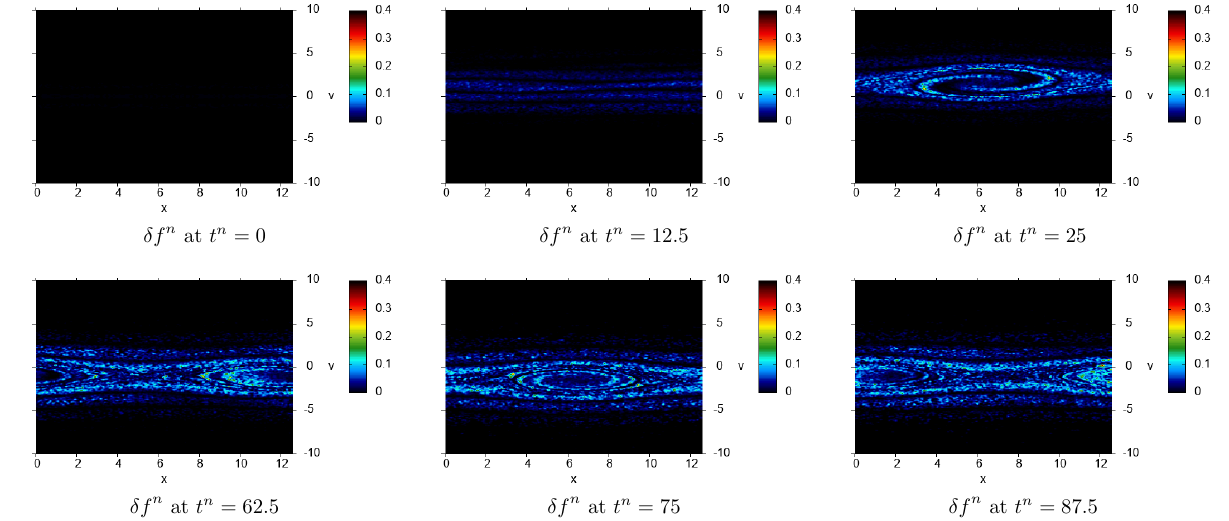}
  \caption{
  Snapshots of the residual density $\df^n$ corresponding to the FBL-$\df$ PIC simulation shown in 
  Figure~\ref{fig:DTSI_FBL_df_evol_f}.
  }
  \label{fig:DTSI_FBL_df_evol_df}
\end{figure*}

In this section we assess the accuracy and denoising capabilities of the basic $\df$ PIC and FBL-$\df$ PIC methods
by plotting various densities in phase space. In Figures~\ref{fig:TSI_f}--\ref{fig:DTSI_FBL_df_evol_df} 
the axes show $x \in [0,L]$ and $v \in [-v_{\rm max},v_{\rm max}]$ with $L$ and $v_{\rm max}$ as indicated above,
and the same color range is used for the purpose of comparison.

In Figures~\ref{fig:TSI_f} and \ref{fig:DTSI_f}, we first plot the densities 
$f^n$ corresponding to the different methods listed in Table~\ref{tab:methods}
for both the TSI and DTSI test cases, at the time $t^n = 100$ where the exact solutions
coincide.

For both test cases we observe very noisy (oscillating) profiles for the coarse full-$f$ PIC solutions
which use about 20,000 particles, and smoother profiles for the fine PIC solutions
which use 16 times as many particles. FBL solutions are essentially free of spurious oscillations,
with many fine structures well resolved in the fine solutions compared to the coarse ones.
We also note that the TSI and DTSI solutions are in very good agreement at $t^n = 100$ for these 
full-$f$ numerical simulations.

Regarding the $\df$ solutions, we observe that in the TSI test case (Fig.~\ref{fig:TSI_f}) both the basic $\df$ PIC
and the FBL-$\df$ PIC methods allow to strongly reduce the level of oscillations compared to the coarse
PIC simulation -- all three using approximatively 20,000 'active' particles. In the DTSI case however 
(Fig.~\ref{fig:DTSI_f}), we see a significant reduction in the accuracy of the basic $\df$ solution,
where the global central vortex suffers from a visible deterioration and spurious streams of particles have appeared at high-velocities.
In contrast, the accuracy of the FBL-$\df$ PIC method in the DTSI case remains very comparable to that of the TSI case.
In particular we observe that the new method is able to capture several low-density filaments, 
i.e., fine regions with zero or very low density distinct from the central vortex. 
In the basic $\df$ simulation these fine structures are poorly resolved, as we can see
from Fig.~\ref{fig:DTSI_f} where the low-density filaments on the left of the central vortex are virtually merged with the latter.
To better visualize the difference in the two methods, we next show in Figures~\ref{fig:DTSI_df_evol_f} to \ref{fig:DTSI_FBL_df_evol_df} 
a series of plots corresponding to snapshots of various densities at successive times $t^n = 0, 12.5, \dots , 87.5$ for the DTSI test case:
In Figure~\ref{fig:DTSI_df_evol_f} and \ref{fig:DTSI_FBL_df_evol_f} we show the evolution of the $\df$ and FBL-$\df$ solutions,
completed by that of a reference scheme (an FBL simulation using a $1024 \times 1024$ grid) in Figure~\ref{fig:DTSI_ref_evol_f}.
In Figure~\ref{fig:DTSI_FBL_df_evol_*} we then show the evolution of the bulk density $f^n_*$ computed 
by the FBL-$\df$ PIC method, which we remind is periodically remapped on a $50 \times 50$ grid of cubic splines,
see Table~\ref{tab:methods}. Finally the evolution of the residual $\df^n$ densities are shown in Figure~\ref{fig:DTSI_df_evol_df} for the 
basic $\df$ method, and in Figure~\ref{fig:DTSI_FBL_df_evol_df} for the FBL-$\df$ method.

From these plots we see that the density strongly deviates from its initial value due to the external drive, 
which clearly challenges the basic assumption of a standard $\df$ approach.
Here, we have chosen a deviation so large that the bulk density cannot be compensated by a proper
$\df$ weighting of the particles, leading to significant errors in the full $f^n$ density visible in Figure~\ref{fig:DTSI_df_evol_f}.
As expected the solution computed with the FBL-$\df$ PIC method on Fig.~\ref{fig:DTSI_FBL_df_evol_f} does not show this erratic behaviour.
The main reason for this is visible in Fig.~\ref{fig:DTSI_FBL_df_evol_*} where we see that the coarse bulk density 
computed by the FBL-$\df$ scheme properly follows the total density. As a result the $\df$ particles can effectively represent 
small scale variations with respect to the main part of the plasma density. This is made evident in 
Fig.~\ref{fig:DTSI_df_evol_df} and \ref{fig:DTSI_FBL_df_evol_df} where we see that the amplitude of the 
$\df$ particle density for the FBL-$\df$ scheme is significantly lower than with a basic $\df$ method.

In Fig.~\ref{fig:DTSI_FBL_df_evol_*} we also observe that the numerical bulk density is subject to a visible 
numerical diffusion. This is an expected consequence of the periodic coarse-grid interpolations involved in 
the semi-Lagrangian updates, and over moderate times it should not pose a serious issue since the total density 
is described by the sum of the bulk and the perturbed particle density. Over long simulation times
however, this numerical diffusion will lead to bulk densities that do not approximate well the total $f$. 
In such cases a proper re-initialization of the bulk density $f_*$ will have to be devised, 
based on the available representation of the total density, and various methods do exist in this direction
such as adaptive kernel density estimation algorithms \cite{van_yen_wavelet-based_2010, gao_fft_2017, wu_reducing_2018} 
or adaptive sparse-grid methods \cite{muralikrishnan_sparse_2021}. 
We note that such an algorithm is likely be more costly than a simple update of the coarse bulk density,
but this should be mitigated by the fact that it will be applied on much larger time periods.

\subsection{Density and field plots in physical space}

\begin{figure*} [!htbp]
  \includegraphics[width=\textwidth]{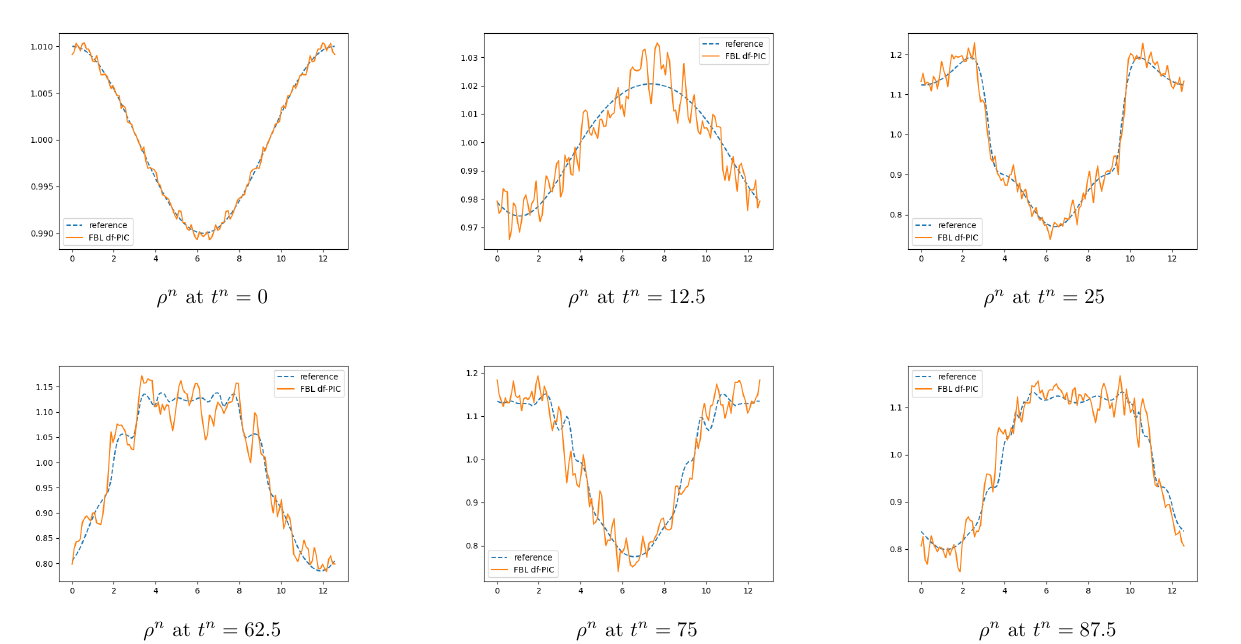}
  \caption{
  Snapshots of the charge density $\rho^n$ deposited with the FBL-$\df$ PIC scheme (with the DTSI test case),
  corresponding to the phase-space densities shown in Figure~\ref{fig:DTSI_FBL_df_evol_f}.
  }
  \label{fig:DTSI_FBL_df_evol_rho}
\end{figure*}

\begin{figure*} [!htbp]
\includegraphics[width=\textwidth]{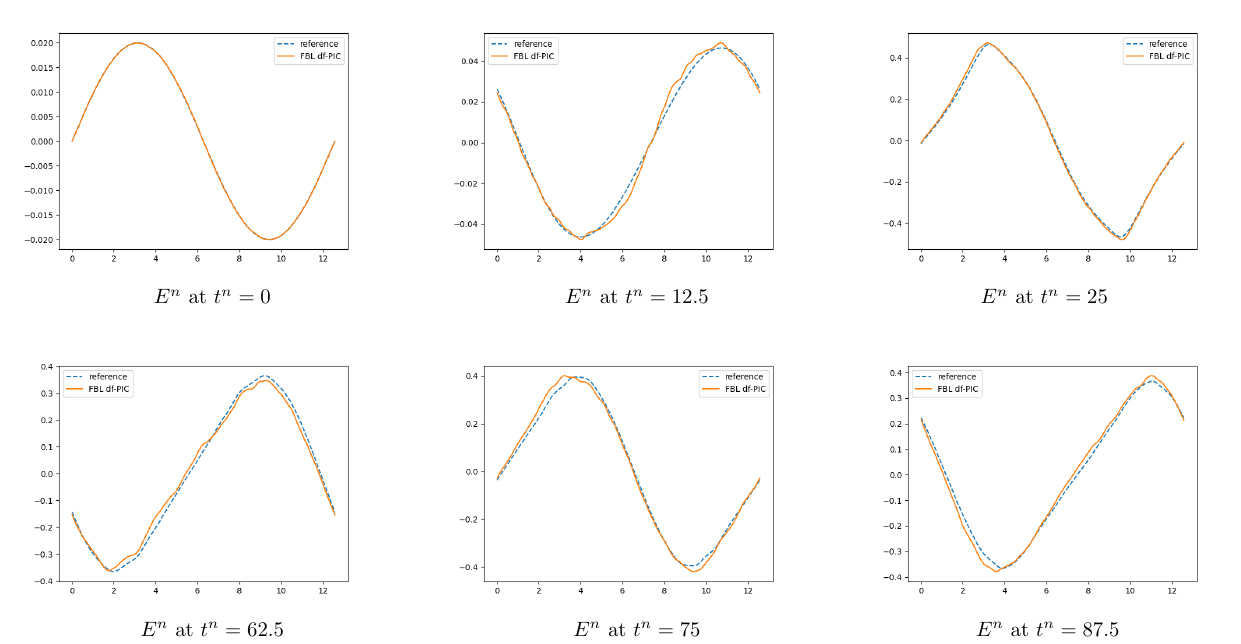}
  \caption{
  Snapshots of the self-consistent field $E^n$ computed by the FBL-$\df$ PIC scheme (DTSI)
  corresponding to the charge densities shown in Figure~\ref{fig:DTSI_FBL_df_evol_rho}.
  }
  \label{fig:DTSI_FBL_df_evol_E}
\end{figure*}

In Figures~\ref{fig:DTSI_df_evol_rho} and \ref{fig:DTSI_FBL_df_evol_rho}
we next plot the snapshots of the deposited charge densities $\rho^n$ 
for the DTSI test case, corresponding to the $\df$ and FBL-$\df$ solutions shown in 
in Figures~\ref{fig:DTSI_df_evol_f} and \ref{fig:DTSI_FBL_df_evol_f}.
For a better assessment of the accuracy, we also plot in dashed line 
a reference charge density computed by a fine semi-Lagrangian scheme.
The associated $E^n$ fields are plotted in Figures~\ref{fig:DTSI_df_evol_E} and \ref{fig:DTSI_FBL_df_evol_E}.

These plots show that despite the velocity averaging, the high level of noise visible in the $\df$ phase-space densities
is also present in the deposited charge densities, which results in significant errors in the self-consistent electric field.

\subsection{Approximation of momentum and energy}

\begin{figure*} [!htbp]
\includegraphics[width=\textwidth]{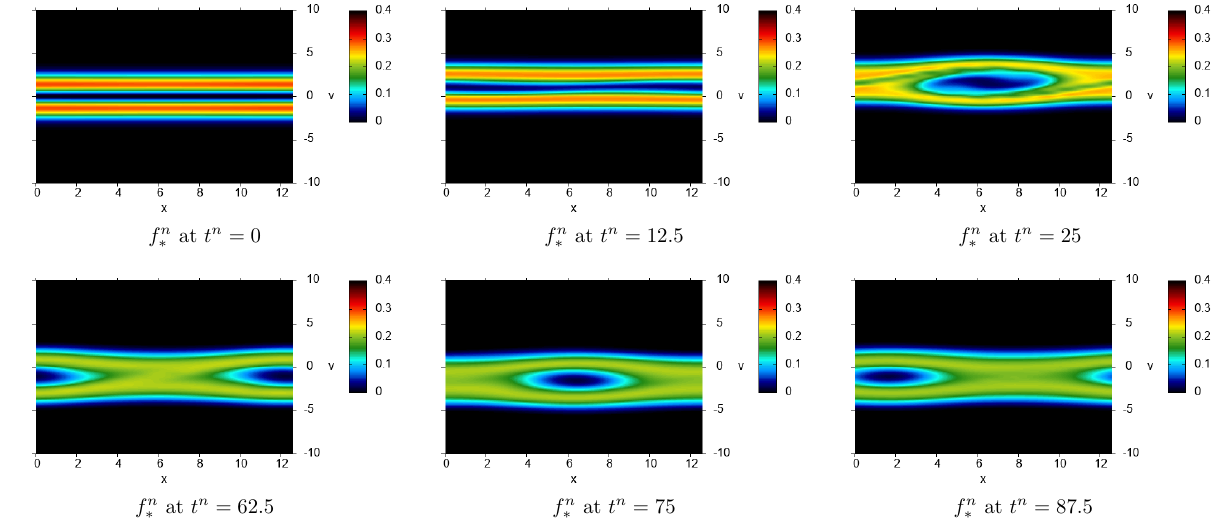}
  \caption{
  Snapshots of the bulk density $f^n_*$ corresponding to the FBL-$\df$ PIC simulation shown in 
  Figure~\ref{fig:DTSI_FBL_df_evol_f}.
  }
  \label{fig:DTSI_FBL_df_evol_*}
\end{figure*}

\begin{figure*} [!htbp]
\includegraphics[width=\textwidth]{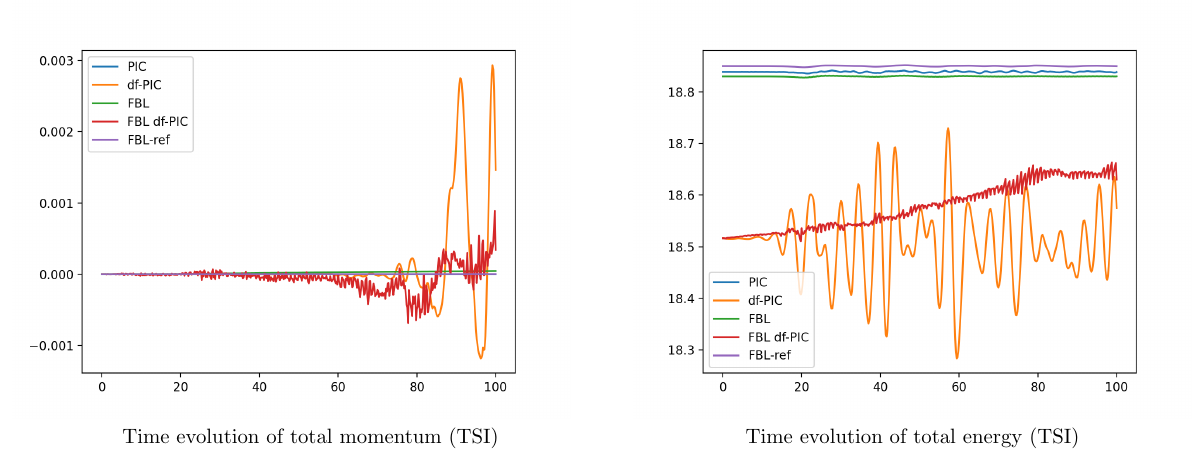}
    \caption{
    Time evolution of the total momentum (left) and energy (right) 
    for the TSI test case.
    }
\label{fig:TSI_energymom}    
\end{figure*}

We next assess the accuracy of the different methods in approximating the system momentum
and energy, namely 
$$
\int_\Omega v f(z) \rmd z
\quad \text{ and } \quad 
\frac 12 \int_\Omega v^2 f(z) \rmd z + \frac 12 \int_0^{L} E(x)^2 \rmd x
$$ 
with $z = (x,v) \in \Omega = [0,L] \times [-v_{\rm max},v_{\rm max}]$ as above. 
In Figure~\ref{fig:TSI_energymom} we plot the time evolution of these discrete quantities
for the TSI test-case, where they are invariants of the exact solution 
(assuming $f = 0$ for $\abs{v} > v_{\rm max}$).
The first finding is that both the full-$f$ PIC and the FBL scheme 
(for the coarse and fine meshes) preserve these quantities 
with very good accuracy, while visible errors accumulate for the two $\df$ methods.
More precisely, we observe oscillating errors for both $\df$ schemes, with no clear winner: 
although the oscillations are clearly of larger amplitude with the basic $\df$ scheme,
a secular growth is visible on the numerical energy for the FBL-$\df$ method,
which may be due to the numerical diffusion of the bulk density (visible in 
Fig.~\ref{fig:DTSI_FBL_df_evol_*}) caused by the periodic coarse updates as discussed above.
We note however that for the TSI test-case both $\df$ schemes did also show similar 
accuracy in the density plots of Fig.~\ref{fig:TSI_f}, so that their similar 
performance in the conservation properties may not come as a big surprise.

\begin{figure*} [h]
\includegraphics[width=\textwidth]{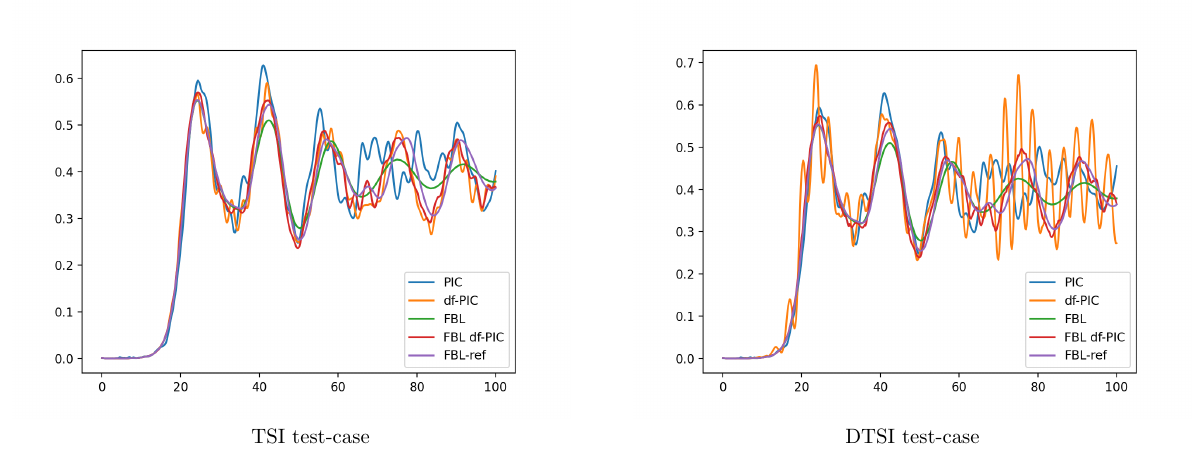}
    \caption{
      Time evolution of the self-consistent electric energy 
      for the TSI (left) and DTSI (right) test cases.
    }
\label{fig:E_energy}    
\end{figure*}

For the DTSI test-case the total momentum and energy are no longer system invariants 
due to the external drive. To compare the performances of the different methods
we thus plot in Figure~\ref{fig:E_energy} the energy of the self-consistent
electric field, for both the TSI (left) and DTSI (right) test-cases. 
Here the first observation is that the results present a high level of variability 
for the different methods. However, not all methods perform equally. 
If we consider that in both test-cases the fine FBL method (purple curve) 
can be used as reference, then we find that the curves of the FBL-$\df$ scheme (in red)
are of reasonable accuracy. More precisely they are less damped than those of the 
coarse FBL scheme (in green), and they show less oscillations than those of the coarse full-$f$
PIC scheme (in blue). In the TSI test-case the curve of the basic $\df$ scheme (in orange) shows
a similar behaviour with slightly more oscillations, while in the DTSI test case the oscillations
are of very high amplitude (close to the actual level of the energy itself). 
As a result, we thus find that for the DTSI test-case the new scheme also shows a superior accuracy 
in terms of energy approximation.

\subsection{Reduction of the statistical errors}

We conclude our numerical experiments by considering the empirical variances 
for the total particle number and current, as defined by Equations (4.80) and (4.81)
in Ref.~\cite{KHHMBBS_2019}, namely
\begin{equation} \label{sigma_N}
    \sigma_N^2 = \frac{N_p^2}{N_p-1} \Big[\sum_{k = 1}^{N_p} (\delta w_k)^2 
        - \Big(\sum_{k = 1}^{N_p} \delta w_k\Big)^2 \Big]
\end{equation}
and 
\begin{equation} \label{sigma_J}
    \sigma_J^2 = \frac{N_p^2}{N_p-1} \Big[\sum_{k = 1}^{N_p} (v_k \delta w_k)^2 
        - \Big(\sum_{k = 1}^{N_p} v_k \delta w_k\Big)^2 \Big]
\end{equation}
respectively. We note that here the weight normalisation 
differs from that of Ref.~\cite{KHHMBBS_2019}: 
for $\df$ and full-$f$ simulations (where the weights are constant) we have  
$$
\delta w^n_k \approx \frac{1}{N_p}\frac{\df^n(z_k)}{g^n(z_k)}
\qquad \text{ and } \qquad 
w^n_k = w^0_k = \frac{1}{N_p}\frac{f^0(z_k)}{g^0(z_k)},
$$
respectively, see \eqref{dw} and \eqref{w}. 
In particular $\sigma_N = 0$ for full-$f$ runs using a sampling  
distribution $g^0 = f^0 / (\int f^0(z) \rmd z)$.

Following the interpretation of the $\df$ method as a variance reduction technique \cite{Aydemir_1994}, 
these respective quantities may be used as quantitative indicators of the statistical errors associated 
with the Monte Carlo approximation of the total density $\iint f(t,x,v) \rmd x \rmd v$ 
and normalized current $\iint f(t,x,v) v \rmd x \rmd v$, see Ref.~\cite[Sec.~2]{KHHMBBS_2019}.

\begin{figure*} [!htbp]
\includegraphics[width=\textwidth]{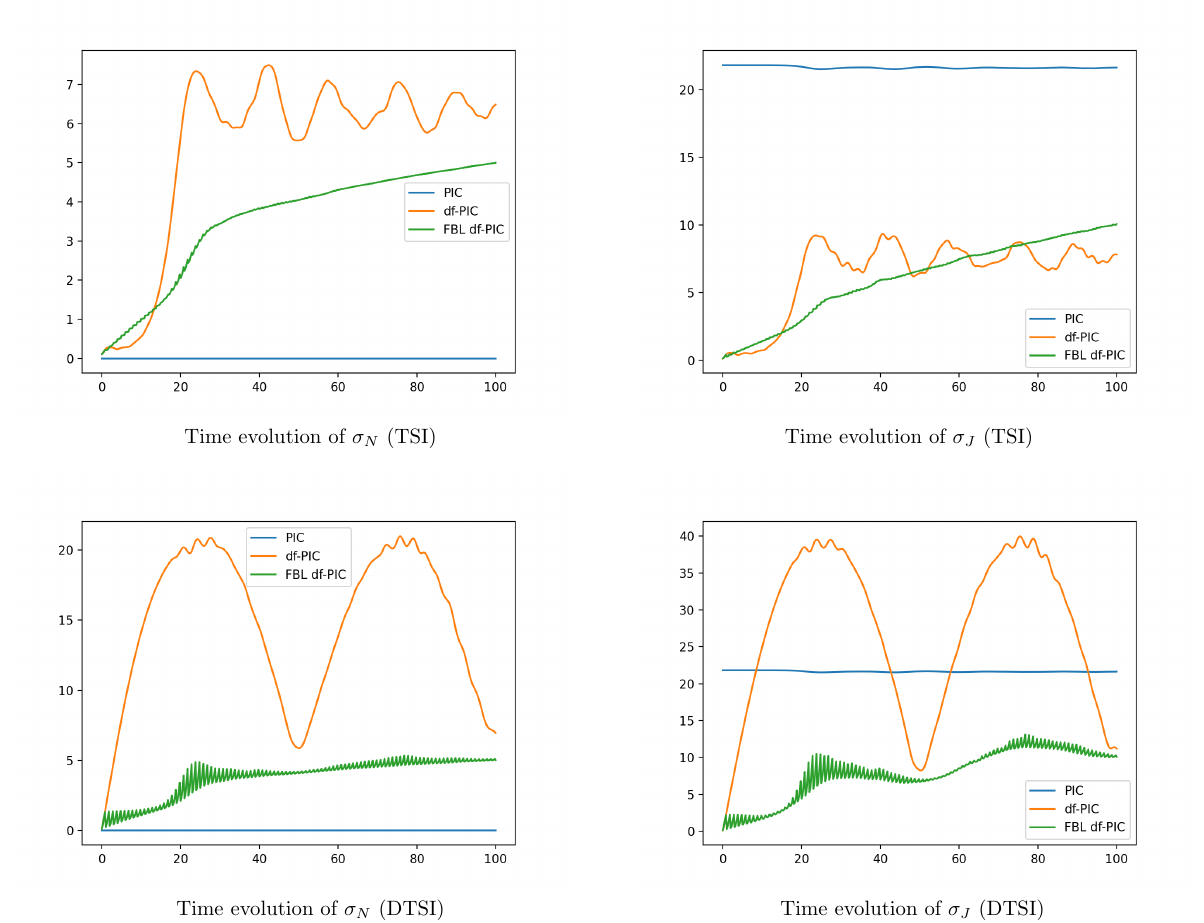}
  \caption{
  Time evolution of the empirical variances for the TSI (top row) and DTSI (bottom row) test cases.
  }    
\label{fig:TSI_sigma}    
\end{figure*}

In Figure~\ref{fig:TSI_sigma}, we plot these quantities for the two test cases, 
and for the three coarse PIC schemes,
namely the coarse full-$f$ PIC, the basic $\df$ and the FBL-$\df$ scheme.
For the TSI test case corresponding to the top plots, we observe that the $\df$ schemes indeed 
result in a reduction of the current variance $\sigma_J$, of a factor between two and three 
(the full-$f$ density variance is zero as previously noticed).
We also note a secular growth in the case of the FBL-$\df$ scheme: 
this is probably due to the slow diffusion of the bulk density caused by the periodic remappings,
and should be mitigated by proper re-initialization methods for the bulk density.
We refer to the perspective section below for a discussion on possible improvements.

\begin{figure*} [!htbp]
\includegraphics[width=\textwidth]{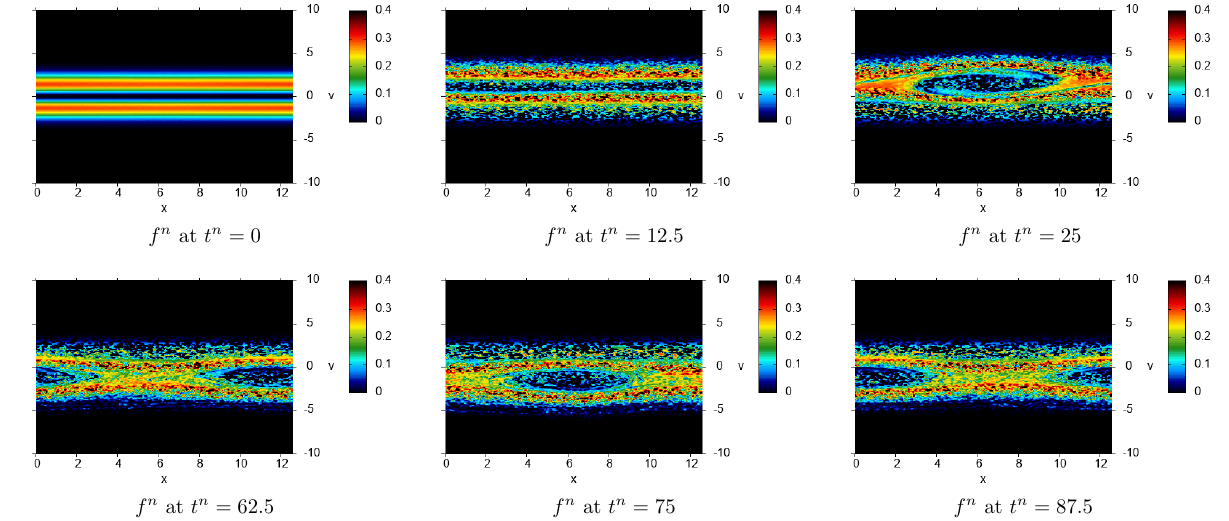}
\caption{  
  Snapshots of the density $f^n$ computed by the $\df$ PIC scheme at various times $t^n$, for the DTSI test case.
  At time $t^n = 100$, the corresponding density is shown in the upper-right plot of Fig.~\ref{fig:DTSI_f}.
  }
  \label{fig:DTSI_df_evol_f}
\end{figure*}

\begin{figure*} [!htbp]
\includegraphics[width=\textwidth]{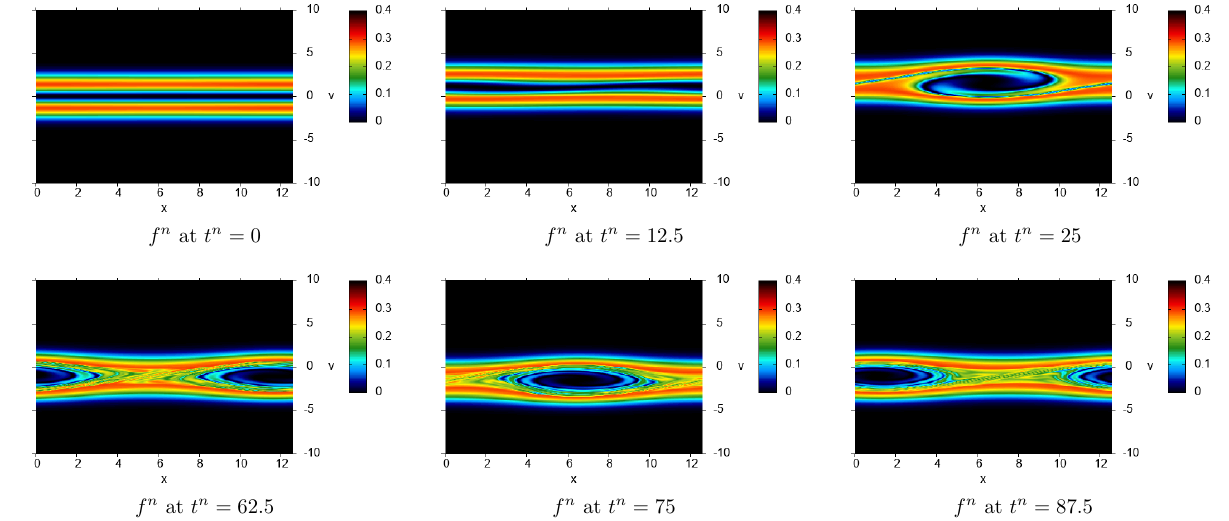}  
  \caption{
  Snapshots of the density $f^n$ obtained with a reference method 
  (an FBL simulation using a $1024 \times 1024$ grid)
  for the same times than in Figures \ref{fig:DTSI_df_evol_f} and \ref{fig:DTSI_FBL_df_evol_f}, 
  again for the DTSI test case.
  }
  \label{fig:DTSI_ref_evol_f}
\end{figure*}

\begin{figure*} [!htbp]
\includegraphics[width=\textwidth]{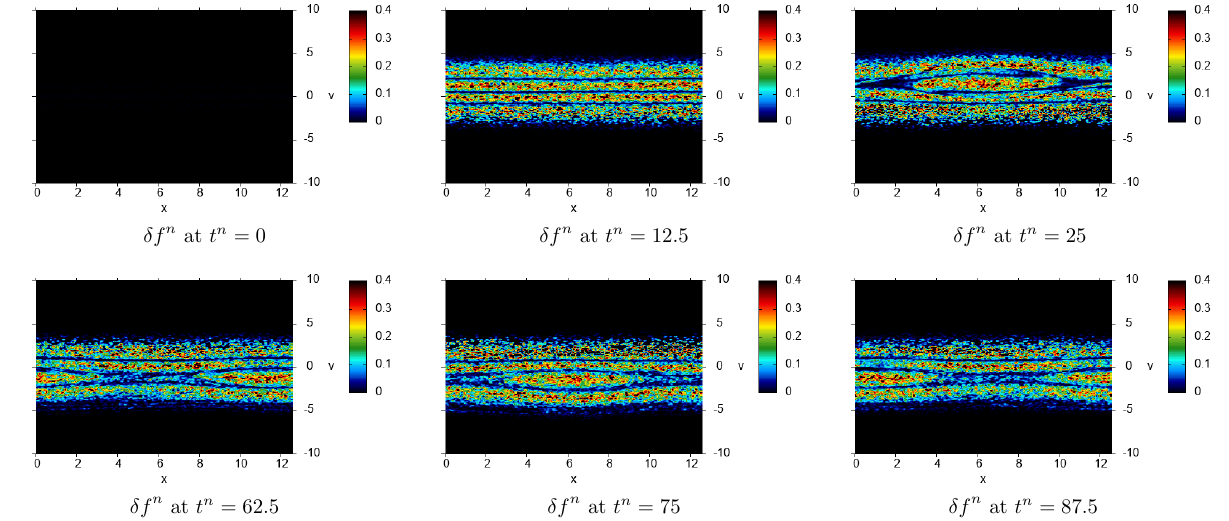}
  \caption{
  Snapshots of the residual density $\df^n$ corresponding to the $\df$ PIC simulation shown in 
  Figure~\ref{fig:DTSI_df_evol_f}.
  }
  \label{fig:DTSI_df_evol_df}
\end{figure*}

\begin{figure*} [!htbp]
\includegraphics[width=\textwidth]{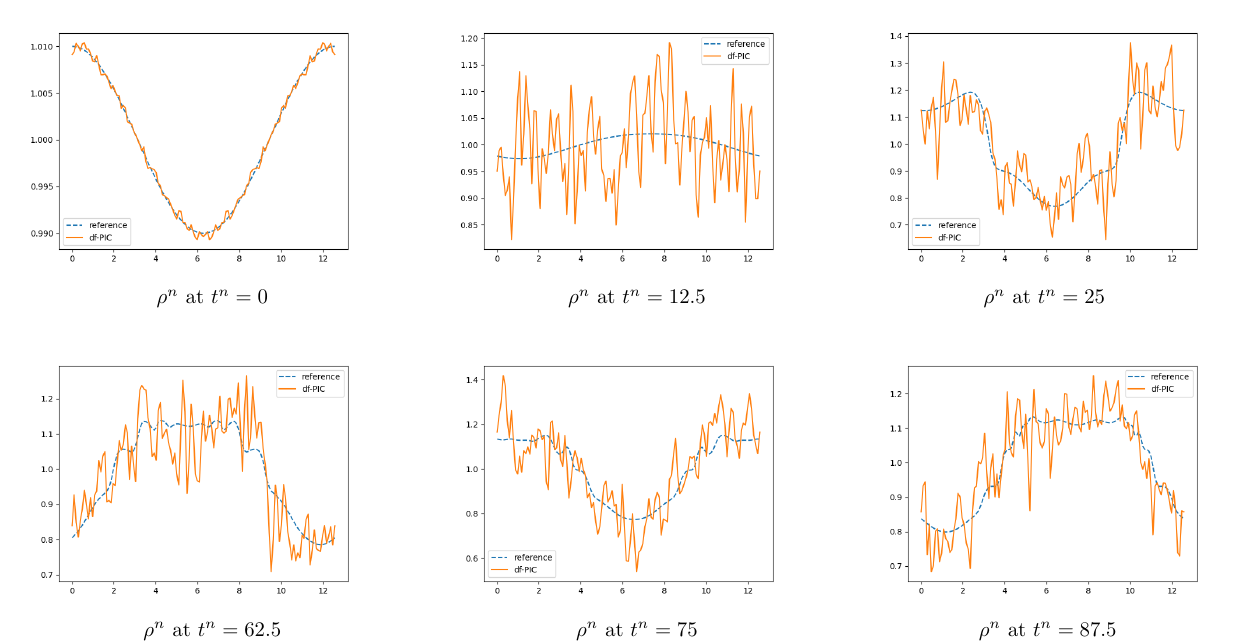}
  \caption{
  Snapshots of the charge density $\rho^n$ deposited with the $\df$ PIC scheme (with the DTSI test case),
  corresponding to the phase-space densities shown in Figure~\ref{fig:DTSI_df_evol_f}.
  }
  \label{fig:DTSI_df_evol_rho}
\end{figure*}

\begin{figure*} [!htbp]
\includegraphics[width=\textwidth]{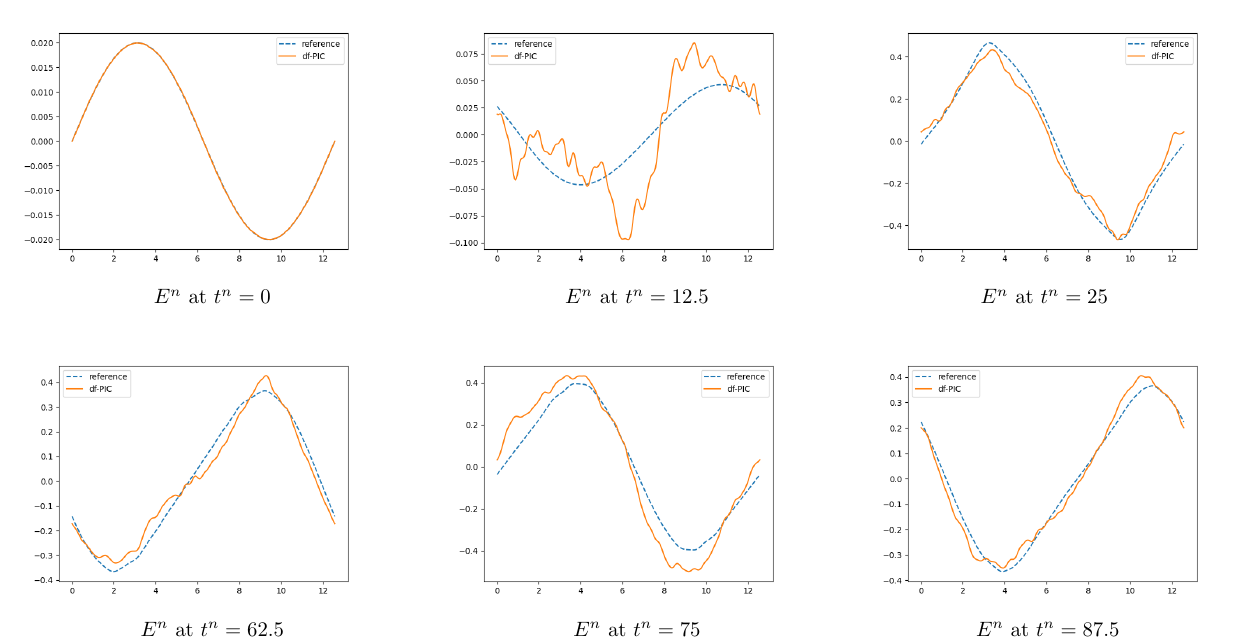}
  \caption{
  Snapshots of the self-consistent field $E^n$ computed by the $\df$ PIC scheme (DTSI),
  corresponding to the charge densities shown in Figure~\ref{fig:DTSI_df_evol_rho}.
  }
  \label{fig:DTSI_df_evol_E}
\end{figure*}

For the DTSI test case, we observe similar values for the current variances $\sigma_J$
of the full-$f$ FBL-$\df$ PIC simulations (with values around 20 and 10, respectively). 
In contrast, strong oscillations are visible in the variances of the basic $\df$ PIC scheme, 
which are closely correlated with the deviation in velocity already seen in Figure~\ref{fig:DTSI_df_evol_f}, 
and the associated offset with the initial density. The amplitude of these oscillations (around 40)
confirm our previous observations that a basic $\df$ approach is no longer justified when the 
density deviates significantly from its initial profile, and further validates the proposed 
method in such regimes.

\section{Conclusion}
\label{sec:conclu}

\subsection{Summary}

In this article we have presented a $\df$ PIC method where the bulk density is
represented by smooth B-splines on a coarse grid, and periodically updated with a forward-backward Lagrangian (FBL) method.
This method amounts in updating the spline coefficients of the bulk density with a backward semi-Lagrangian method,
where the backward trajectories are obtained by local inversions of the forward trajectories provided by the PIC code.
In these inversions the local backward flow is represented as a quadratic mapping whose Jacobian and Hessian matrices
are computed from the relative displacements of regular passive markers pushed forward by the PIC code.
After describing the representation of the bulk and variation density in terms of weighted collections of shape functions
with variable width, and recalling the general form of full-$f$ and $\df$ particle approximations to transport problems, we have
provided a detailed description of a remapped $\df$ scheme where the bulk density updates are performed using the FBL method.
Finally we have studied the basic performances of the method on two academic examples involving one-dimensional 
(i.e., 1D1V phase-space) electrostatic problems:
a simple two stream instability where the deviation from the initial density is mild and does not justify any update
of the bulk density, and a driven one where an oscillating electric field induces strong displacements of the phase-space
density in velocity. In this latter case the basic equilibrium / deviation separation assumption is no longer valid and the $\df$
method fails as expected to correctly resolve the solution. In contrast, the FBL-$\df$ method shows good results, and a measurable
reduction of the statistical error.

\subsection{Novelty of the proposed approach}

Although we did not compare the numerical performances of our method with the
direct projection scheme of Ref.~\cite{Allfrey_Hatzky_2003}, we believe that opting
for semi-Lagrangian updates of the bulk density is a significant novelty
which has the key feature that it is does not rely on an accurate particle approximation
of the total density, but rather on the accurate particle {\em trajectories}.
Two typical configurations may illustrate why this difference is meaningful: the first one is
a regime where a localised plasma blob moves through an otherwise empty phase-space.
In a region crossed by the blob, the bulk density should ideally go from zero to some
positive density, and then go back to zero as the blob leaves the region.
With a direct projection scheme however this cannot happen: as the $\df$ particles
enter the region the bulk density will be updated by their coarse projection and take a positive value,
but as they leave they will not be able to exactly compensate the smooth bulk density
(and neither will their projection), because they have precisely {\em moved away} from it.
As a result the updated bulk density will retain a spurious residual of that fine discrepancy,
which will remain there since the $\df$ particles have now left.
Another problematic scenario is the opposite one, where both the density and the transport flow
are smooth in phase space. If the bulk density grid is not too coarse, then one will always find
some phase-space cells devoid of markers: on such cells the projected $\df$ density will be zero
by construction, leading to strong oscillations.
Because the forward-backward Lagrangian approach relies on the particle trajectories and
a smooth grid-based bulk density, it is exempt of both these pitfalls: For the case of a moving blob
this has already be seen on the DTSI test-case where the external drive moves the plasma density
over large portions of the phase-space, with Fig.~\ref{fig:DTSI_FBL_df_evol_*}
showing a clean transport of the bulk density (e.g. between times $t^n = 0$ and $25$).
For the second one the key point is that a smooth flow leads to accurate backward trajectories,
according to the analysis of Ref.~\cite{Campos-Pinto.Charles.2018.jcm}. On every grid node the new value
will thus correspond to that of the previous bulk density at a point where the
smooth spline density should be also accurate, independent of the markers resolution.
\medskip

We also point out that this approach not only allows to approximate general bulk densities, but it can be applied
to other representations, as long as their evaluation is easily computable:
instead of B-splines, one could for instance use piecewise polynomials or a collection of Fourier modes.
The semi-Lagrangian updates would then involve specific algorithms (such as interpolation or FFT) for updating the weights,
based on nodal evaluations of the previous density at the feet of the backward characteristics computed by the FBL scheme.

\subsection{Perspectives}

Moving forward, an important direction of research is to extend the present method to more challenging problems,
such as higher dimensional ones. We are in particular interested in gyrokinetic models where the transport
phase-space has typically 4 dimensions. In their current form the various algorithms (namely, the FBL flow
reconstruction and the spline quasi-interpolation) can be directly applied to a 4D phase space,
however preliminary numerical simulations have shown that typical transport flows in higher dimensions 
have an anisotropic smoothness which is not well captured by our isotropic algorithms.
Our plan is thus to design extensions of the current approach that will be able to efficiently reconstruct
high-dimensional characteristic flows with anisotropic smoothness.

\medskip

Another improvement will be to design good re-initialization schemes for the bulk density that could be applied 
on larger time scales to counter the numerical diffusion inherent to the coarse semi-Lagrangian updates,
as well as the secular growths in the energy error and in the variance, observed in Fig.~\ref{fig:TSI_energymom} and \ref{fig:TSI_sigma}.
Adaptive approximation methods do exist
which have already shown good results for particle denoising,
such as adaptive wavelet or Fourier filters \cite{van_yen_wavelet-based_2010, gao_fft_2017}
kernel density estimation algorithms \cite{wu_reducing_2018} which automatically select the optimal width for the particle shape functions,
or sparse-grid adaptive methods \cite{muralikrishnan_sparse_2021} which have also proven effective for particle denoising in higher
dimensions.

% 
% \section*{Data Availability Statement}
% 
% The code used to produce the results presented in this article is available on request from the authors.

\section*{Acknowledgments}
The authors thank Roman Hatzky for fruitful discussions on statistical error reduction, 
as well as the anonymous reviewers for their comments which have improved the presentation of our results.
This work has been carried out within the framework of the EUROfusion Consortium and has received funding from the Euratom research and training programme 2014-2018 and 2019-2020 under grant agreement No 633053. The views and opinions expressed herein do not necessarily reflect those of the European Commission.
% \end{acknowledgments}

\nocite{*}
\bibliographystyle{plain} %apsrev4-1}
\bibliography{fbl_df_pic}

\end{document}